\documentclass[pre,a4paper,twocolumn,showpacs,showkeys,nofootinbib,preprintnumbers,amsmath,amssymb,10pt]{revtex4}

\usepackage{graphicx}
\usepackage{comment}
\usepackage{amssymb,amsmath,amsthm,enumitem}
\usepackage{epstopdf}
\usepackage{color}
\usepackage[top=2.1cm, bottom=2cm,left=1.5cm, right=1.5cm,headheight=2cm,headsep=0.7cm]{geometry}
\usepackage{float}

\begin{document}

\title{Locality of interactions for planar memristive circuits}
\author{F. Caravelli}
\affiliation{Theoretical Division and Center for Nonlinear Studies,\\
Los Alamos National Laboratory, Los Alamos, New Mexico 87545, USA }

\begin{abstract}
The dynamics of purely memristive circuits has been shown to depend on a projection operator which expresses the Kirchhoff constraints, is naturally non-local in nature and does represent the interaction between memristors. In the present paper we show that for the case of planar circuits, for which a meaningful Hamming distance  can be defined, the elements of such projector can be bounded by exponentially decreasing functions of the distance. We provide a geometrical interpretation of the projector elements in terms of determinants of Dirichlet Laplacian of the dual circuit.
 For the case of linearized dynamics of the circuit for which a solution is known, this can be shown to provide a light cone bound for the interaction between memristors.
This result establishes a finite speed of propagation of signals across the network, despite the non-local nature of the system. 
\end{abstract}

\maketitle

\section{Introduction} 
The study of memristors and their nonlinear properties has become an area of interest both for experimentalists \cite{indiveri,Avizienis,Stieg12} and theorists \cite{chua76a,pershin11a,Caravelli2015,Caravelli2016, Caravelli2016rl}. Their employment in circuits has been shown to lead to non-trivial dynamics and computation abilities \cite{traversa14b,Traversa2014,diventra13a} which are currently under careful scrutiny. Memristors are passive components that can be thought of as a time varying resistance, and which depends on an internal state. These have been studied for several years, and their main characteristic is a pinched hysteresis loop in the Current-Voltage diagram when controlled in alternate current \cite{chua76a, strukov08, stru12}. More recently, Resistive Random Access Memory (RRAM) devices have been shown to further generalize this type of behavior \cite{Valov}. The collective dynamics of memristors has been however only recently being tackled theoretically \cite{Caravelli2015,Stieg12} and, despite the effort, very little is known about their interaction. Albeit the characterization of a single memristor is rather simple from a modelling perspective, when these components become part a network the Kirchhoff constraints introduce an extra layer of nonlinearity which is yet poorly understood. A better characterization of this type of non-linearity might provide a theoretical understanding of crossbar arrays technology \cite{Lu} and their limitations.
Moreover, it is not clear how the non-locality introduced by the constraints affects the memristor's behavior, as it has been  already pointed out \cite{Sheldon, Lu2, Lu3}. Understanding the role of non-locality is thus an outstanding problem.
This paper provides evidence that that the strength of interaction is bounded exponentially in the distance between components, and provide a geometrical interpretation to the interaction matrix between memristors in terms of the Dirichlet Laplacian. The Dirichlet Laplacian is a discrete Laplacian in which Dirichlet boundary conditions are enforced.
As an application, we show that temporal correlations are bounded by an effective speed of propagation of information, result which emerges naturally as the exponential bound is established.

Memristors are interesting as general purpose computing devices since these can be used to store, propagate and process information in the same device. 
Physical memristors are not perfectly state preserving, and slowly relax to the limiting resistance even when a voltage is not applied \cite{AtomicSwitch1,AtomicSwitch2}. 
The internal memory parameter $w$ of a TiO2 or Ag$\ ^+$ single memristor (which can be thought of as the charge in the conductor) is driven by an external voltage source $S$, and evolves according to the equation $$\partial_t w(t)=\alpha w(t)-\frac{R_{on}}{ \beta } I = \alpha w(t)-\frac{ R_{on}}{ \beta } \frac{S}{R(w)},$$
where $0\leq w(t)\leq1$ is the internal memory parameter of the memristor, $ R(w)= R_{on} (1-w)+R_{off} w$ is the resistance. In this parametrization,  $R_{on}$ and $R_{off}$ are the limiting resistances for $w=0$ and $w=1$ respectively, and $\alpha$ and $\beta$ are constants which set the timescales for the relaxation and excitation of the memristor respectively. 

Recently, a differential equation for the internal memory of a purely memristive circuit was obtained in \cite{Caravelli2016rl}. Such an equation is first order, autonomous and incorporates all the circuit constraints.  The single memristor equation can in fact be generalized to a coupled non-linear differential equation for memristors on a network:
\begin{eqnarray}
 \frac{d \vec W(t)}{d t} &=&\alpha \vec W(t)- \frac{1}{\beta}( I+\xi\ \Omega  W(t))^{-1} \Omega \vec { S}(t),
\label{eq:dynnn5}
\end{eqnarray}
where $ I$ the identity matrix, $\vec S$ is the vector of sources in series to each memristor, $\xi=\frac{R_{off}-R_{on}}{R_{on}}$ and $\vec W(t)$ is a vector which contains the internal memory parameter for each memristor, and $W(t)$ is a diagonal matrix composed of the elements of $\vec W(t)$. For a purely memristive circuit (e.g. only passive memristor and voltage sources being present), eqn. (\ref{eq:dynnn5}) describes the evolution of the internal memory consistently, as it satisfies all the circuit constraints at each instant of time. Moreover, shows the separation between the topology of the circuit from the internal memory of each memristor.  Importantly enough eqn. (\ref{eq:dynnn5}) exhibits several symmetries, and has been related to self-organizing maps and machine learning \cite{Caravelli2016ml,Prezioso,Sheridan}.
A prominent role is in fact played by the matrix $ \Omega$ which encodes the circuit topology, and is a projector on the loop space of the circuit.\footnote{For details of this formalism and a derivation, we point out the Appendix of \cite{Caravelli2016rl} or \cite{Caravelli2016ml}. Also, we note that the notation and nomenclature is different from the graph theory literature. What here we intend with \textit{loops}, in graph theory are commonly called \textit{cycles} (formed of a set of distinct edges).} If $ \Omega$ were to be diagonal, each memristor would behave independently from the other. Thus, the  sparsity (or denseness) of this matrix characterizes the (non-)locality of the interaction. 

At the origin of the bound presented in this paper there is the numerical evidence obtained in \cite{Caravelli2016rl}, where it was observed that $\Omega$ has elements which fall off exponentially with a Hamming distance on the graph for a planar circuit. We provide a precise statement of the numerical result mentioned above, which is however based on the Hamming distance on the \textit{dual} graph, i.e. the one between the loops.
Our result is in some sense of general applicability, as it provides bounds on the projection operator based on the topological features of planar graphs. Our result has thus also applications in graph theory. If $B$ is the directed incidence matrix of a graph and $A$ its directed loop matrix, since $\Omega\equiv \Omega_A=I-B^t(BB^t)^{-1}B=I-\Omega_{B^t}$ due to the duality $A^t B=B^tA=0$, any result regarding $\Omega$ can be inferred also for $\Omega_{B^t}$.

\section{Planar circuits: a geometric interpretation for  $\Omega$.} The proof of the exponential bound can be logically split into  three major steps: \textit{a)} the identification of a formula for $\Omega$ in terms of Gram matrices on the dual graph; \textit{b)} The interpretation of Gram matrices in terms of a Dirichlet Laplacian; \textit{c)} deriving an exact bound for each term of the expansion.

\subsection{Reduction of the projection operator by Gram-Schmidt decomposition}
The necessary trick which will allow to infer the bounds is based on the fact that
any projector $\Omega_A=A(A^t A)^{-1}A^t$ can be written as $\Omega_A=\tilde A\tilde A^t$ where $\tilde A$ is composed of the orthonormalized  rows of the matrix $A$. In doing so we trade the complexity of the inverse with the complexity induced by the orthogonalization process. We first recall the definition of the matrix $A$. Each (fundamental) loop in the circuit  is composed of only $\pm1, 0$ and constructed as follows. First, we assign an orientation to each edge of the graph representing the circuit, and an orientation to a loop.  If a loop $l$ does not contain an edge $\beta$, then $A_{\beta l}=0$. If the orientation of the loop agrees with the orientation of the edge we assign $+1$, and $-1$ otherwise. It is important to note that set of fundamental loops is a basis which spans the loop space of the dual circuit, and thus any Gram determinant on these vectors will be necessarily non-zero. These vectors compose the matrix $A$, whose rows we can thus write as $\vec A_{l}$. 
Using the formula $\Omega=\tilde A \tilde A^t$ we now have a way to characterize the operator $\Omega_{ij}$ in a geometric fashion using the Gram-Schmidt decomposition. The rows of the matrix $\tilde A $ represent (orthonormalized fundamental) loops, meanwhile the columns represent edges of the graphs. 
Each column of $\tilde A$ has been orthonormalized such that $\tilde A^t \tilde A=I_L$, where $L$ is the total number of loops and $I_L$ is the identity matrix of size $L\times L$. The first step of the Gram-Schmidt decomposition is the introduction of  matrix $A^\prime$ whose rows are an \textit{orthogonal set}. This can be written as:
\begin{equation}
\vec A^\prime_{l}=\vec A_{l}-\sum_{j=1}^{l-1} \frac{\langle \vec A^\prime_{j},\vec A_{l}\rangle }{\langle \vec A^\prime_{j},\vec A^\prime_{j}\rangle} \vec A^\prime_{j}.
\end{equation}
The key trick which will allow us to prove the exponential bound for the elements of $\Omega_{ij}$ is based on writing the Gram-Schmidt decomposition in a non-recursive manner (\cite{Gantmacher}, pp.  256-258). This is based on the following expression in terms of a determinant:
\begin{eqnarray}
\tilde A_p=\text{det}\left(
\begin{array}{ccccc}
\langle \vec A_1, \vec A_1 \rangle & \langle \vec A_1, \vec A_2 \rangle & \cdots & \langle \vec A_1, \vec A_{p-1} \rangle & \vec A_1\\
\langle \vec A_2, \vec A_1 \rangle & \langle \vec A_2, \vec A_2 \rangle & \cdots & \langle \vec A_2,  \vec A_{p-1} \rangle & \vec A_2 \\
\vdots & \vdots & \ddots & \vdots  & \vdots\\
\langle \vec A_p, \vec A_1 \rangle & \langle \vec A_p, \vec A_2 \rangle & \cdots & \langle \vec A_p, \vec A_{p-1} \rangle &  \vec A_p
\end{array}
\right).
\end{eqnarray}
Despite the unusual meaning to be associated to this determinant (it represents a vector and not a scalar), it can be written as the following sum of vectors by following the usual rules for the co-factor expansion:
\begin{eqnarray}
\tilde A_p=(-1)^p\sum_{k=1}^p (-1)^{k} \frac{G_p^k}{\sqrt{G_{p} G_{p-1}}} \vec A_k,
\end{eqnarray}
where $G_0=1$, and $G_p$ is the Gram determinant of the loop vector space, i.e. the determinant of the matrix:
\begin{eqnarray}
G_p&=&\text{det}\left( 
\begin{array}{ccc}
\langle \vec A_1, \vec A_1 \rangle  & \cdots & \langle \vec A_1, \vec A_p \rangle \\
\vdots  & \ddots & \vdots \\
\langle \vec A_p, \vec A_1 \rangle &  \cdots & \langle \vec A_p, \vec A_p \rangle 
\end{array}
\right) =\text{det}({\tilde {\mathcal L}}_p),\nonumber
\end{eqnarray}
where we have introduced the definition of the matrix ${\tilde {\mathcal L}}_p$, meanwhile $G^k_p$ is the determinant of the matrix 
\begin{eqnarray}
G_p^k&=&\text{det}\left( 
\begin{array}{ccc}
\langle \vec A_1, \vec A_1 \rangle &  \cdots & \langle \vec A_1, \vec A_{p-1} \rangle \\
\vdots & \ddots   & \vdots \\
\langle \vec A_p, \vec A_1 \rangle &  \cdots & \langle \vec A_p, \vec A_{p-1} \rangle 
\end{array}
\right) =\text{det}({\tilde {\mathcal L}}_p^k),\nonumber
\end{eqnarray}
where the $k$-th row and $p$-th column have been removed from the matrix ${\tilde {\mathcal L}}_p$.\footnote{Thus, ${\tilde {\mathcal L}}_p$ is a matrix of of size $p\times p$ and ${\tilde {\mathcal L}}_p^k$ is of size $(p-1)\times(p-1)$.}\\
\subsection{Gram matrices for the loop space of planar graphs}  
\begin{figure}[ht!]
\centering
\includegraphics[scale=0.2]{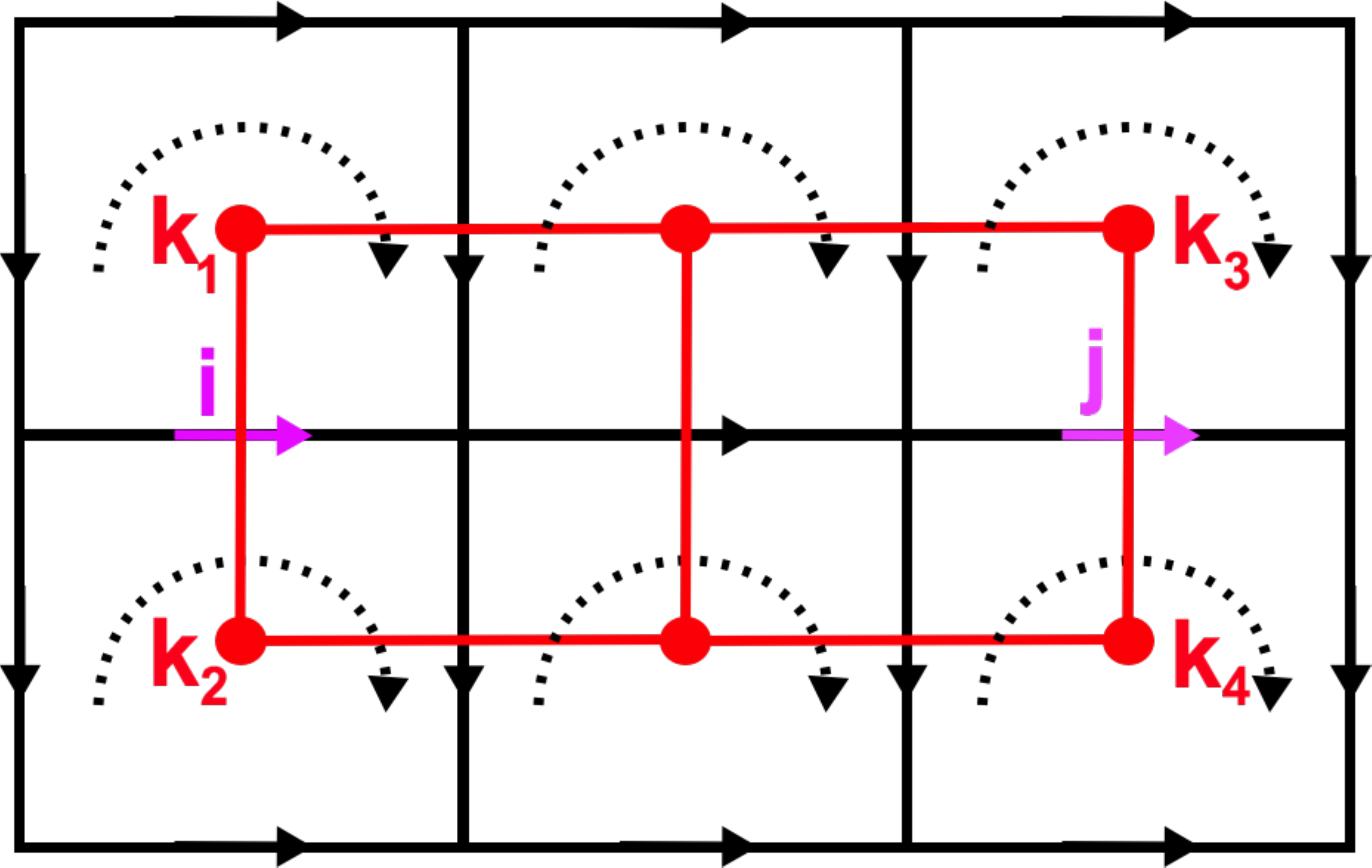}
\caption{The adjacency matrix on the loop dual space is associated with the red graph, whereas the black lines are associated to the circuit. Each loop for planar graphs can be oriented alike. }
\label{fig:orientation}
\end{figure}

The previous considerations are of general character. We now specialize to the case of planar circuits.
It is important to note that, if we introduce the adjacency matrix $M_{ij}$ associated with the representation of the dual graph (the loops), we have the identity:
$$\langle \vec A_i,\vec A_j\rangle=\begin{cases}- M_{ij} & \textit{if }i\neq j, \\ |\mathcal L_i| & \textit{if }i=j, \end{cases}$$
where $|\mathcal L_i|$ is the length of the loop $i$.
This identification applies to planar graphs.\footnote{It is likely that this relation can be generalized to some non-planar graphs, but in this paper we focus on the simplest case.}
In order to see this, let us first note that the diagonal elements correspond to the square of the elements on the loop, and since the elements of $\vec A_{i}=\{\pm 1,0\}$, we have as a result that $\langle \vec A_l,\vec A_l\rangle=\sum_i (\vec A_l)_i^2\equiv |\mathcal L_l|$ is the length of the loop $l$. This result is independent from the orientation either of the loops or of the edges of the graph as it can be easily observed. In order to understand why the off-diagonal elements of the matrix $R_p$ can be instead associated to the adjacency matrix, we rely on Fig. \ref{fig:orientation} to help us visualize this fact.
Let us for instance focus on loops $k_1$ and $k_2$. We assume that these loops have in common a single edge (memristor) $i$. If the two loops are oriented alike (which occurs for planar graphs), necessarily one has $A_{k_1 i}A_{k_2 i}=-1$. If we sum over the index $i$, for general graphs the scalar product $\langle \vec A_{k_1}, \vec A_{k_2}\rangle=-z$, where $z$ is the number of edges which these loops have in common.\footnote{If the graph is not planar, $\langle A_i,A_j\rangle=\epsilon_{ij} M_{ij}$, where $\epsilon_{ij}$ can take values $\pm 1$.} In the case in which each loop has at most one edge in common to another loop, then the off diagonal elements can be associated to the adjacency matrix of the dual graph which connects two adjacent loops, and $z=1$.\footnote{We avoid here for simplicity the case of a more general structure such as a \textit{multigraph}. In general, if the dual graph is more complex and the loops have more edges in common, the dual graph can be a multigraph with multiple edges between two nodes.}

\begin{figure}[b!]
\centering
\includegraphics[scale=0.28]{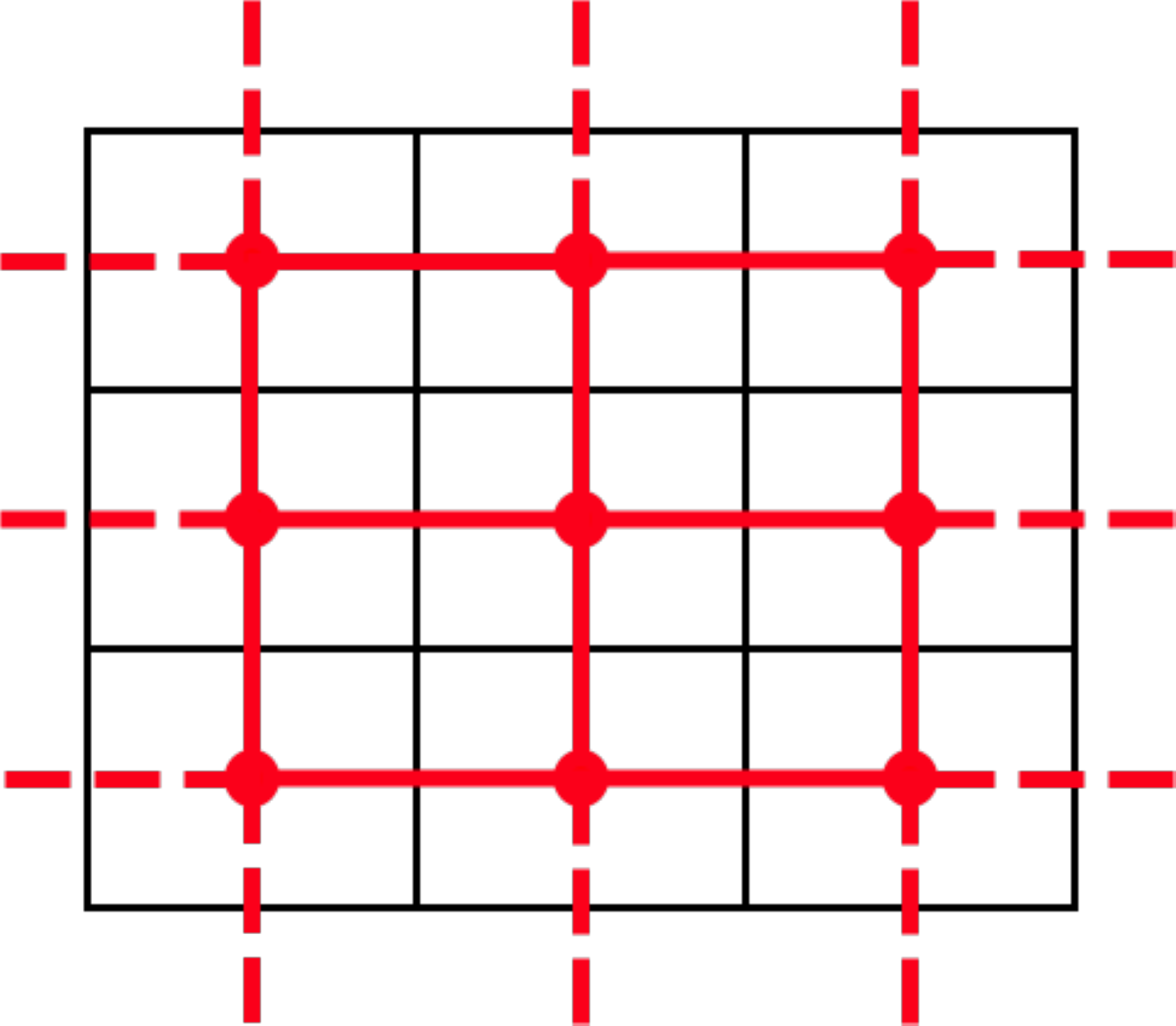}
\caption{The regularizer of the laplacian operator $\partial L$ is associated with the half edges (dashed lines) in the graph above, which are edges not crossed by any edge in the dual graph representation. Full edges are those associated with the adjacency matrix $M_{ij}$. This shows that the determinantal equations are related to the eigenvalues of the Dirichlet Laplacian on dual the graph.}
\label{fig:boundary}
\end{figure}

 Albeit the matrix ${\tilde {\mathcal L}}_p$ is diagonally dominant, it is not strictly diagonally dominant and thus it can have, in principle, null eigenvalues. This is important, as the matrix ${\tilde {\mathcal L}}_p$ resembles the discrete Laplacian of the dual graph, which is known to contain a number of zero modes equal to the number of connected components. However, we aim to show that ${\tilde {\mathcal L}}_p$ is a regularized Dirichlet Laplacian instead.
 For this purpose, we can try to further understand the matrix ${\tilde {\mathcal L}}_p$ by noticing that the discrete operator $\tilde {\mathcal L_p}$, such that $G_p=\text{det}(\tilde {\mathcal L}_p)$ can effectively be written as $\tilde {\mathcal L_p}=\mathcal L_p+\partial { \mathcal L_p}$, where ${\partial {\mathcal L}}_p$ is a diagonal matrix which we now define and $\mathcal L_p$ is the Laplacian of the dual graph. The Laplacian of the dual graph is defined as $D-M$, where $D$ is the degree matrix of the dual graph, which is defined as $D_i=\sum_{j}M_{ij}$, and $M$ is its adjacency matrix. In the bulk of the dual graph, the operator $\tilde {\mathcal L}$ does coincide with the Laplacian matrix. However, at the boundary it does not. We note that $\tilde {\mathcal L}_p$ differs from the Laplacian matrix on the diagonal only. In fact, $( {\partial \mathcal L_p})_{ii}=D_{ii}-|\mathcal L_i|$ and represents the difference between the length of the loop $|\mathcal L_i|$ and the degree matrix $D$ which is associated with the connectivity of the dual graph. In order to understand why $\partial {\mathcal L}_p$ has support only on the boundary of the dual graph, let us visually inspect $\tilde {\mathcal L}_p$ in a simple example, as in Fig. \ref{fig:boundary}. The half edges which are drawn (dashed lines) are associated with the memristors which are present on the boundary of the circuit. $( {\partial {\mathcal L_p}})_{ii}$ is thus equal to the number of half edges attached to node $i$, which are memristors not crossed by any edge in the dual graph. Since these are non-zero only on the boundary, $\partial {\mathcal L}_p$ can be interpreted as a discrete operator with support only on the boundary. We note then that the diagonal elements of  $\partial {\mathcal L}_p$ count the edges which leave the boundary. Thus $\tilde {\mathcal L}_p$ is the definition of the so called (graph) Dirichlet Laplacian (of a subgraph of size $p$), which is known to have only positive eigenvalues \cite{Chung}. This confirms the fact that the determinants $G_p$ and $G_p^k$ are meaningful and non-zero, and provides an interesting geometrical interpretation to the matrix $\tilde{ \mathcal L_p}$.\\
\subsection{The bound}
We now focus on calculating the projection operator. For the first vector, the Gram-Schmidt procedure gives $\tilde A_1=\frac{\vec A_1^\prime}{||\vec A_1^\prime||}$.
The matrix $\tilde A_{pi}$ can be thus be written as 
$ \tilde A_{pi}=(-1)^p\sum_{k=1}^p (-1)^{k} \frac{G_p^k}{\sqrt{G_{p} G_{p-1}}}  (\vec A_k)_i$.
From the geometric point of view, each loop vector is associated with the dual of the circuit. The quantities $\langle \vec A_{j},\vec A_{l}\rangle$ are non-zero if and only if the loops $j$ and $l$ are adjacent in the dual graph.
Giving the preamble on the geometrical meaning of the determinants, the projection operator can be written by squaring the matrix $\tilde A$ by summing over the loop indices:
\begin{eqnarray}
\Omega_{ij}&=&\sum_{l=1}^L \tilde A_{li} \tilde A_{lj}=\sum_{l=1}^L (-1)^{2l}\sum_{k=1}^l (-1)^{k} \frac{G_l^k}{\sqrt{G_{l} G_{l-1}}}  (\vec A_{k})_i  \nonumber \\
&\cdot& \sum_{k^\prime=1}^l (-1)^{k^\prime} \frac{G_l^{k^\prime}}{\sqrt{G_{l} G_{l-1}}}  (\vec A_{k^\prime})_j\nonumber \\
&=&\sum_{l=1}^L \sum_{k, k^\prime=1}^l (-1)^{k+k^\prime} \frac{G_l^k G_l^{k^\prime}}{G_{l} G_{l-1}}  (\vec A_{k})_i    (\vec A_{k^\prime})_j
\end{eqnarray}
which is valid for arbitrary graphs. We now note that we can introduce the elements of the inverse of the matrix of $R_p$ by noticing that $(R_p^{-1})_{ij}=(-1)^{i+j}\frac{\text{det}(R_i^j)}{\text{det}(R_p)}=(-1)^{i+j}\frac{G_i^j}{G_p}$, from which we obtain
\begin{eqnarray}
\Omega_{ij}&=&\sum_{l=1}^L \sum_{k, k^\prime=1}^l  \frac{G_{l}}{G_{l-1}} \frac{ (-1)^{k+k^\prime} G_l^k G_l^{k^\prime}}{G_{l} G_{l} }  (\vec A_{k})_i    (\vec A_{k^\prime})_j \nonumber \\
&=&\sum_{l=1}^L \sum_{k, k^\prime=1}^l  \frac{G_{l}}{G_{l-1}} \frac{ (-1)^{k+l}  G_l^k  }{G_{l}} \frac{ (-1)^{k^\prime+l}G_l^{k^\prime}}{G_{l} } (\vec A_{k})_i    (\vec A_{k^\prime})_j \nonumber \\
&=&\sum_{l=1}^L \sum_{k, k^\prime=1}^l  \frac{G_{l}}{G_{l-1}} ({\tilde {\mathcal L}}_{l}^{-1})_{lk}({\tilde {\mathcal L}}_{l}^{-1})_{lk^\prime} (\vec A_{k})_i    (\vec A_{k^\prime})_j .
\end{eqnarray}
We now use the fact that ${\tilde {\mathcal L}}_l$ are bounded functions of the adjacency matrix $M$ of the dual graph. 
 In this case, we can write the following bound for ${\tilde {\mathcal L}}_l^{-1}$ \cite{Bounded} based on the graph distance $d(i,j)$ on the \textit{dual graph}:
\begin{equation}
({\tilde {\mathcal L}}_l)_{ij}^{-1}\leq \kappa (X) K \lambda^{d(i,j)}=e^{-z d(i,j) +\rho},
\end{equation}
where $\kappa (X)$ is the Kreiss constant of the matrix $X$ of the eigenvectors of the adjacency matrix $M$\footnote{The Kreiss constant $\kappa(X)$ is defined as  $$\kappa(X)=\text{sup}_{|z|>1} (|z|-1) ||(z I-X)^{-1}||,$$ with $||\cdot||$ the sup norm of a matrix.}, while $K$ is a positive constant and $0<\lambda<1$. Also, we defined for simplicity $z=-\log(\lambda)$ and $\rho=\log(\kappa (X) K)$. Since ${\tilde {\mathcal L}}_l$ is a positive submatrix, we can bound such sum using the full matrix ${\tilde {\mathcal L}}_L\equiv{\tilde {\mathcal L}}$ by extending the sum:
\begin{eqnarray}
\Omega_{ij}&\leq &\sum_{l=1}^L \sum_{k,k^\prime=1}^L  \frac{G_{L}}{G_{L-1}} {\tilde {\mathcal L}}^{-1}_{lk}{\tilde {\mathcal L}}_{l k^\prime}^{-1} (\vec A_{k})_i    (\vec A_{k^\prime})_j,
\end{eqnarray}
which can be further bounded by:
\begin{eqnarray}
\Omega_{ij}&\leq & \sum_{l=1}^L \sum_{k, k^\prime=1}^L  \frac{G_{l}}{G_{l-1}} ({\tilde {\mathcal L}}^{-1})_{lk}({\tilde {\mathcal L}}^{-1})_{lk^\prime} (\vec A_{k})_i    (\vec A_{k^\prime})_j.
\end{eqnarray}
We can thus write the following bound:
\begin{eqnarray}
\Omega_{ij}&\leq & \sum_{k, k^\prime=1}^L  \sum_{l=1}^L \frac{G_{l}}{G_{l-1}} ({\tilde {\mathcal L}}^{-1})_{lk}({\tilde {\mathcal L}}^{-1})_{lk^\prime} (\vec A_{k})_i    (\vec A_{k^\prime})_j \nonumber \\
&\leq& \sum_{k, k^\prime=1}^L  \sum_{l=1}^L \frac{G_{l}}{G_{l-1}} e^{-z d(l,k) +\rho}e^{-z d(l,k^\prime) +\rho} (\vec A_{k})_i    (\vec A_{k^\prime})_j   .\nonumber 
\label{eq:bound1}
\end{eqnarray}
By using the triangle inequality, $d(x,y)\leq d(x,z)+d(z,y)$ for any point $z$, we can further simplify the expression and introduce the upper bound
$e^{-z(d(l,k)+d(l,k^\prime)}\leq e^{-z d(k,k\prime)}$.
In order to obtain the final bound we need to understand the factor $\frac{G_l}{G_{l-1}}$. 
As we have mentioned before, $R_p$ is a positive matrix. We note that for any positive semi-definite matrix of the form $Q=\left(A\ B^t;\ B\ C\right)$, where $A$ and $C$ are square matrices and $B$ is rectangular, then we have $\text{det}(Q)\leq \text{det}(A) \text{det}(C)$. This statement can be proven easily by means of Schur complements.\footnote{In fact, using the Schur complement we see that $\text{det}(Q)=\text{det}(A) \text{det}(C-B^t A^{-1} B))$. It is a property of the Schur complement of positive matrices that $C-B^t A^{-1} B$  and $B^t A^{-1} B$ are positive matrices. Now for positive definite symmetric matrices one has $\text{det}(A+B)\geq \text{det}(A)+\text{det}(B)$. Then we have $\text{det}(C)=\text{det}(C-X+X)\geq \det{det}(C-X)+\text{det}(X)\geq\text{det}(C-X)$. From which the result follows.} Let us choose $Q$ to be ${\tilde {\mathcal L}}$ and $A={\tilde {\mathcal L}}_{p-1}$, which are indeed both semi-positive matrices. For this choice, $C$ is a constant, and thus $\det(C)=({\tilde {\mathcal L}}_p)_{pp}$. Then we obtain the bound $\frac{G_{p}}{G_{p-1}}=\frac{\text{det}({\tilde {\mathcal L}}_p)}{\text{det}({\tilde {\mathcal L}}_{p-1})}\leq ({\tilde {\mathcal L}}_{p})_{pp}$. We can thus bound this quantity by the maximum loop length in the circuit, which we define as $\tilde l$. We now notice that the right hand side of eqn. (\ref{eq:bound1}) does not depend anymore on the index $l$.
We can thus write the final bound:
\begin{equation}
\Omega_{ij}\leq L \tilde l \sum_{k,k^\prime=1}^L   e^{-z d(i,j)+\rho} (\vec A_k)_i (\vec A_k)_j.
\end{equation}
Considering that $(\vec A_k)_i$ can be either positive, negative or zero, we can bound absolute value of this quantity by the number of loops each memristor is adjacent to. In the case of Fig. \ref{fig:orientation}, this number is $2$ because of the planarity requirement. We  thus have:
\begin{equation}
|\Omega_{ij}|\leq 4\tilde l L  e^{-z d(i,j)+\rho}=e^{-z d(i,j)+\tilde \rho},
\label{eq:finalresult}
\end{equation}
which the main result of this paper, for positive quantities $L$, $\tilde l$, $\xi$ and $\tilde \rho=\rho+\log\left(4\tilde l L\right)$. 
This bound has been observed numerically in \cite{Caravelli2016rl}, where however we had introduced a similar but different notion of Hamming distance which is however only a linear transformation of the one we used. Eqn. (\ref{eq:finalresult}) shows that the non-locality due to the constraints is bounded exponentially in the distance between the memristors.

\begin{figure}[ht!]
\centering
\includegraphics[scale=0.2]{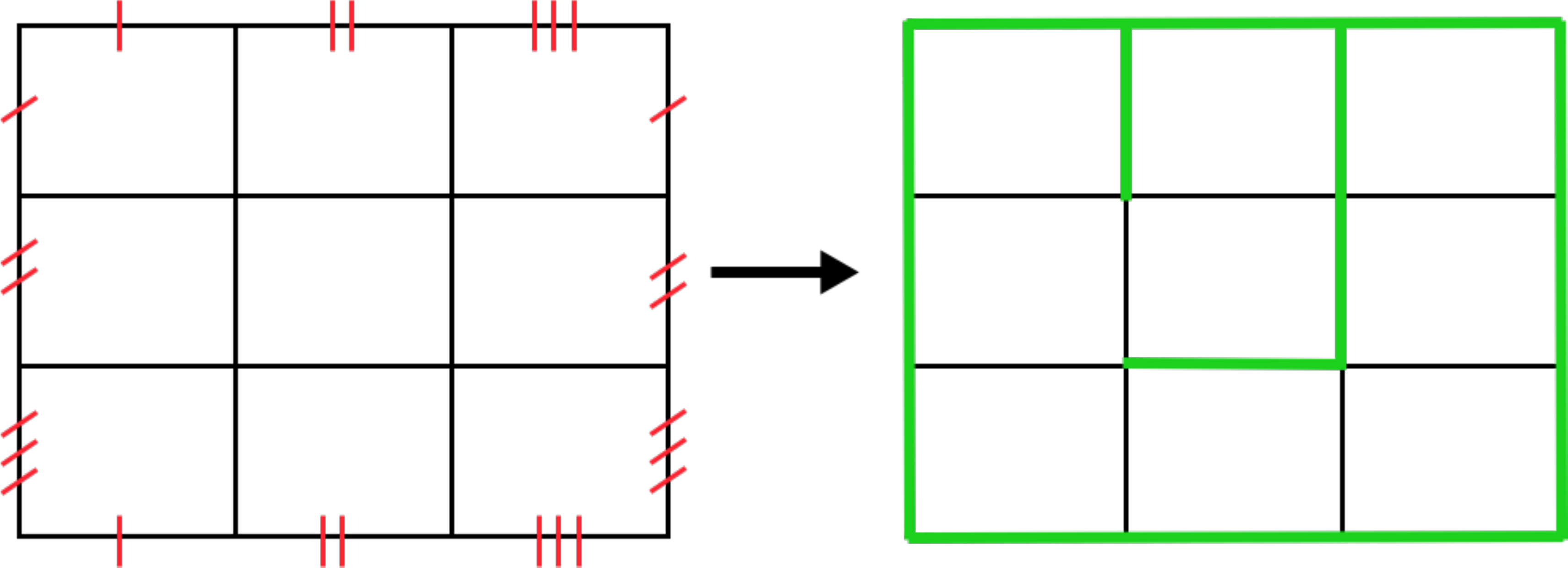}
\caption{Number of independent loops in the case of a circuit without boundary, a torus. The green edges are those associated to a maximal spanning tree $\mathcal T$ for the specific case of a torus.}
\label{fig:torus}
\end{figure}
\subsection{Case of planar surfaces with non-trivial genus} When the surface  representing the circuit does not have any boundary, the operator ${\tilde {\mathcal L}}_l$ should be exactly the Laplacian matrix for the dual graph. In this case, the Laplacian has always a zero mode which would imply $G_l=0\ \forall l$. 
In fact we aim to show the following: $G_l=\det{{\tilde {\mathcal L}}_l}\neq0$ for any circuit $G$. What impedes for boundary-less graphs the emergence of a null eigenvalue? We analyze the graph of Fig. \ref{fig:torus}, which represents a torus. In this case, the number of independent loops associated to the circuit depends on the cardinality of the co-tree $\bar {\mathcal T}$, which is linear $g$, the genus of the surface in which the circuit is embedded. To see this, for any graph embedded in a planar surface we have the Euler characteristic $\chi=2-2g=N-E+F$,
where $N$ is the number of nodes, $E$ the number of edges and $F$ the total number of loops (faces). If we use the fact that $L=E-|\mathcal T|-1$, then we obtain $L=F-3-2g$,
which shows that the total number of \textit{fundamental} loops is proportional to both the total number of loops (faces) and the genus of the embedding surface. For Fig. \ref{fig:boundary} and \ref{fig:torus}, $F=12$, but in the latter case the $g=1$. This explains why from the plane to the torus the number of fundamental loops goes from $9$ to $7$ in this case. This effectively introduces a boundary.\\

\section{Bounded correlations} 
Let us now briefly discuss what are the implications of eqn. (\ref{eq:finalresult}) for the time evolution of the memory. We can in fact make some precise statements in the case $\xi\ll1$, for which an exact solution is known for arbitrary times.
As it has been shown in \cite{Caravelli2016rl}, when we neglect non-linearities  we have the following differential equation for the internal memory:
$$\frac{d}{dt} \vec W(t)=\left(\alpha I+\frac{\xi}{\beta} \Omega S\right) \vec W(t)-\frac{\xi}{\beta} \Omega \vec S$$
with $S$ being the diagonal matrix of the constant source vector $\vec S$ .
The exact solution can be written as:
\vspace{1pt}
\begin{equation}
\vec W(t)=e^{t\left(\alpha I+\frac{\xi}{\beta} \Omega S\right)}\vec W_0-\frac{\xi}{\beta}\int_0^t e^{(t-r)\left(\alpha I+\frac{\xi}{\beta} \Omega S\right)}\Omega \vec S\ dr.
\end{equation}
\vspace{1pt}
Let us study in detail now the properties of this solution.
As a first step, let us prove that powers of exponentially bounded matrices are still exponentially bounded. First of all, if $\Omega_{ij}$ satisfies the bound of eqn. (\ref{eq:finalresult}), then also $\Omega_{ij} S$ will, for any finite diagonal matrix $S$. We also have:
\begin{eqnarray}
|(\Omega S)^n_{ij}|&\leq&\sum_{r_1,\cdots, r_n} |\Omega_{ir_1}||S_{r_1}|\cdots|\Omega_{r_n j}||S_{j} |\nonumber \\
&\leq& \sum_{r_1,\cdots, r_n} \text{sup}(S)^n e^{-\xi\left(d(i,r_1)+d(r_1,r_2)\cdots d(r_n,j)\right)+n\tilde \rho} \nonumber \\
&\leq& m^n \text{sup}(S)^n e^{-z d(i,j)+n\tilde \rho} 
\label{eq:bound2}
\end{eqnarray}
where $m$ is the total number of memristors, and from which we obtain
\begin{eqnarray}
|(e^{\frac{\xi}{\beta} t \Omega S})_{ij}|\leq e^{-z \left(d(i,j) - \frac{m e^{\tilde \rho} \frac{\xi}{\beta} \text{sup}(S)}{z} t\right)}
\end{eqnarray}
which shows, after a brief calculation and introducing $v=\frac{m e^{\tilde \rho} \frac{\xi}{\beta} \text{sup}(S)}{z}$, that
\begin{eqnarray}
|W_i (t)-W_j(0)|&\leq&  
  \sum_j \left(e^{-z(d(i,j)-v t)+\alpha \delta_{ij}}-\delta_{ij}\right)W_j(0) \nonumber \\
&+& \frac{\xi}{\beta z v} \sum_k e^{-zd(i,k)+\rho+\alpha \delta_{ik}}\left(1-e^{zv t}\right) S_k  \nonumber,
\end{eqnarray}
which is the final bound. This bound is unfortunately loose, as the speed of propagation depends on the number of memristors.\\
\begin{figure}
\centering
\includegraphics[scale=0.13]{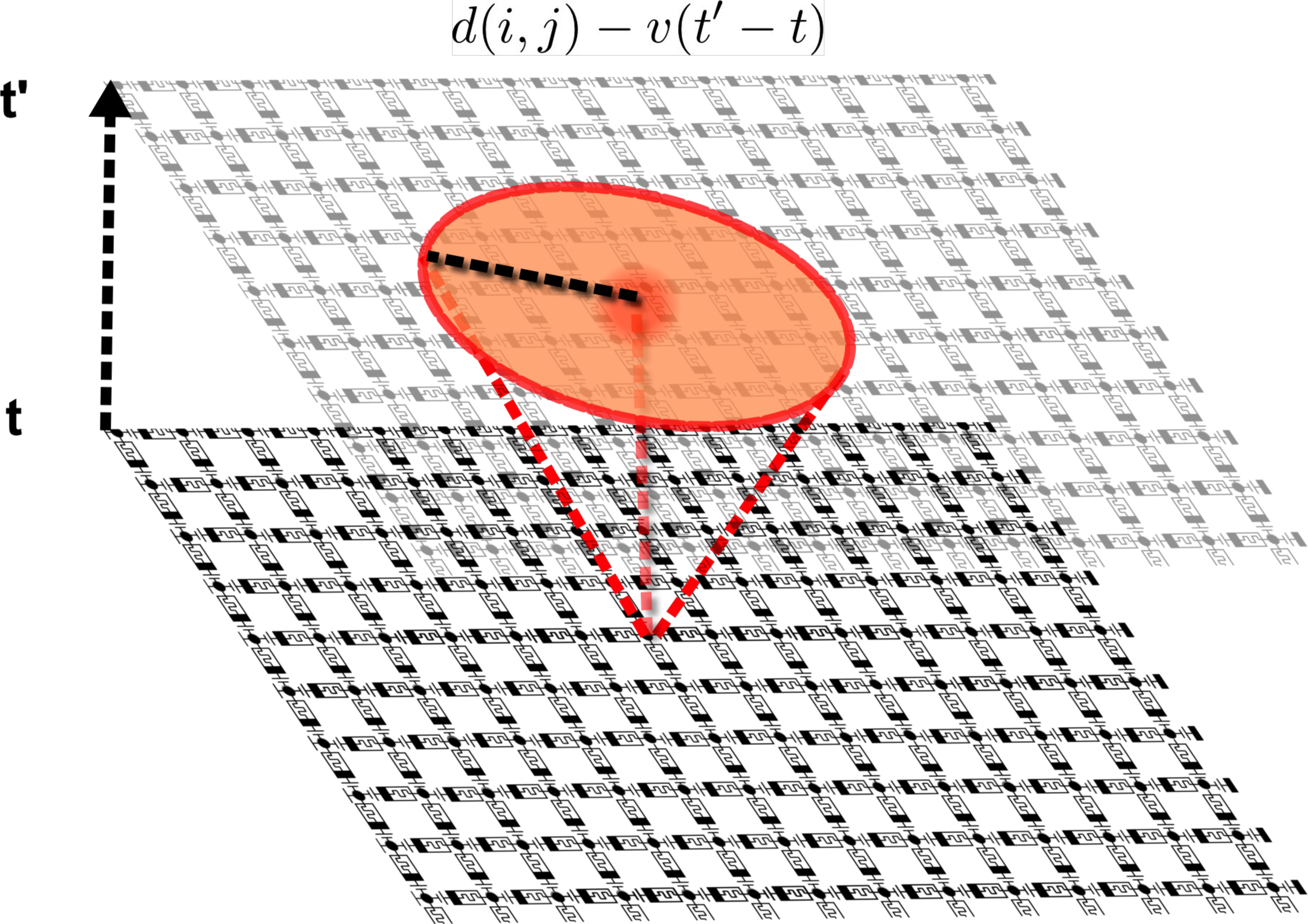}
\caption{Schematic representation of the effective propagation of information. For planar graph correlations fall off exponentially out of an effective light cone with an emergent speed of propagation $v$. Nonlocal effects fall off exponentially outside the cone.}
\label{fig:int}
\end{figure}
The bound of eqn. (\ref{eq:bound2}) can be also applied to $T(\xi)_{ij}=\left((I+\xi \Omega W)^{-1} \Omega\right)_{ij}\leq (1+\xi m e^{\tilde \rho})^{-1} e^{\tilde \rho} e^{-z d(i,j)}$.\\
This shows that, in general, we have the following bound on the derivative \cite{Schuch}:
\begin{equation}
|\frac{d}{dt} W_i|\leq |\alpha W_i+ \frac{e^{\tilde \rho}}{\beta (1+\xi m e^{\tilde \rho})}   \sum_j  e^{-z d(i,j)} S_j|
\end{equation}
which gives the bound:
\begin{eqnarray}
W_i(t)&\leq& e^{\alpha t} W_i(0) \nonumber \\
&-& \frac{e^{\tilde \rho}}{\beta (1+\xi m e^{\tilde \rho})}   \sum_j  e^{-z d(i,j)} \int_0^t e^{-\alpha (t-r)}S_j(r) dr. \nonumber \\
\label{eq:linearbound}
\end{eqnarray}
The bound above shows that the voltage changes affect the memory in an exponential bound in the Hamming distance.

\section{Conclusions} 
Memristive circuits are non-local in nature due to the Kirchhoff constraints. In this paper we have however provided strong evidence that their non-locality is indeed bounded exponentially in the distance between the memristive elements. The proof of this result is based on the understanding of the properties of the projector operator on the space of loops of the circuit, which plays the role of the interaction matrix between the memristors. Our result applies to the case of planar circuits, for which exact properties of the loop vectors can be evinced and connected to the dual graph adjacency matrix, but we believe it might be possible to extend them to a more general class of graphs. Such bound also implies a finite speed of propagation for the interactions in the case of linearized dynamics, $\xi\ll 1$ resembles the general Lieb-Robinson bounds obtained in quantum mechanics or Markov chains \cite{Hasting}. However, the origin of the exponential bound is rather different from the case of a Markov chain. In the present case we have in fact strongly based our computation on the topology and properties of the dual graph, rather than the topology of the circuit. Despite these differences, the meaning to be attributed to them is very similar. 

Our locality result might seem odd at first, as the memristors on a single mesh are non-local in nature. In fact, if all the memristors share the \textit{same} current, then these (whatever the number of memristors) would be strongly correlated.
However, we stress that this result holds for memristors on two different meshes, and that the distance is the one between meshes. We note that a similar eqn. (\ref{eq:dynnn5}) has been derived also for voltage-driven memristors in \cite{Caravelli2016rl}, and thus the result of the present paper applies, to a certain extent, also to other type of memristors.
Further work is necessary to better understand the properties of more complex circuits and tighter bounds on the propagation speed. Yet, this is a step towards the understanding of the propagation of information on circuits with memory. A direct application of our results are bounds on the speed of propagation in crossbar arrays \cite{Kim}. The non-planar case will be the focus of future works.\\

\textit{Acknowledgement.} We would like to thank C. Pagliantini, F.L. Traversa, A. Hamma for various comments on the draft.
We acknowledge the support of NNSA for the U.S. DoE at LANL under Contract No. DE-AC52-06NA25396. 




\begin{thebibliography}{10}

\bibitem{indiveri}
G. Indiveri,    S.-C. Liu,  Proc. of IEEE, 103:(8) 1379-1397 (2015)


\bibitem{Avizienis}
A.V. Avizienis et al.,  PLoS ONE 7(8): e42772. (2012)
 \bibitem{Stieg12} A. Z. Stieg, A. V. Avizienis et al.,  Adv. Mater., 24: 286-293  (2012)

\bibitem{chua76a}
L.~O. Chua, S.~M. Kang.
\newblock { Proc. IEEE}, 64:209--223, 1976.



\bibitem{pershin11a}
Y.~V. Pershin, M. Di~Ventra,  Adv. in Phys., 60:145--227 (2011) 

\bibitem{Caravelli2015}
F. Caravelli, A. Hamma, M. Di Ventra, EPL, 109, 2 (2015)

 \bibitem{Caravelli2016}
F. Caravelli, Front. Robot. AI 3, 18 (2016), arXiv:1511.07135

\bibitem{Caravelli2016rl} F. Caravelli, F. L. Traversa, M. Di Ventra, Phys. Rev. E 95, 022140 (2017)

\bibitem{traversa14b}
F. L. Traversa, M.~Di~Ventra. IEEE Trans. Neural Netw. Learn. Syst. (2015)

\bibitem{Traversa2014}
F. L. Traversa, C. Ramella, F. Bonani, M. Di Ventra, Science Advances, vo. 1, no. 6, pag e1500031 (2015)

\bibitem{diventra13a}
M. {Di Ventra}, Y.~V. Pershin,  Nat. Phys., 9:200 (2013)


\bibitem{strukov08} D.B. Strukov et al., Nature 453, pp. 80-83 (2008)

\bibitem{stru12}
J. J. Yang, D. B. Strukov, D. R. Stewart, Nat. Nano. 8 (2013)



\bibitem{Valov} I. Valov et al, Nature Comm. 4, 1771 (2013)

\bibitem{Lu} W. Lu, Nat. Mat. 12, 93-94 (2013)

\bibitem{Sheldon}
F. C. Sheldon, M. Di Ventra, Phys. Rev. E 95, 012305 (2017)

\bibitem{Lu2} S. Choi, P. Sheridan, W. Lu, Scientific Report 5, 10492 (2015)


\bibitem{Lu3} P. Sheridan, W. Lu, in Memristor Networks, (ed. A. Adamatzy, L. Chua), Springer Int. Pub., Zurich (2014)

\bibitem{AtomicSwitch1}
Ohno et al.,  Appl. Phys. Lett. 99, 203108 (2011);

\bibitem{AtomicSwitch2} Wang et al, Nat. Mat. 16 (2017)


\bibitem{Caravelli2016ml} F. Caravelli,  IJPEDS vo. 0, 1-17 (2017), special issue  Advances in Memristive Networks, ed. Adamatzky et al.

\bibitem{Prezioso} M. Prezioso et al., Nat. Lett. 521, pp 61-62 (2017)

\bibitem{Sheridan} P. M. Sheridan et al., Nat. Techn., doi:10.1038/nnano.2017.83 (2017)

\bibitem{Gantmacher} F. R. Gantmacher, The theory of matrices Vol. 1, Chelsea Publishing Company, New York (1959)

\bibitem{Chung} F. Chung, Spectral Graph Theory, CBMS Lecture Notes, AMS, Providence RI (1997)

\bibitem{Bounded} 
M. Benzi, N. Razouk, Electronic Transactions on Numerical Analysis, V. 28, pp. 16-39, (2007)

\bibitem{Hasting} M.B. Hasting, Phys. Rev. Lett. 93, 140402 (2004)

\bibitem{Kim} S. H. Jo, Nano Lett. 9, 870-874 (2009)



\bibitem{Schuch} N. Schuch et al., 
Phys. Rev. A 84, 032309 (2011)


\end{thebibliography}
\end{document}